\documentclass[aps,twocolumn,showpacs,prl]{revtex4}

\usepackage{dcolumn}
\usepackage{amsmath}
\usepackage{graphicx}

\begin{document}

\title{Generation of microwave radiation in planar spin-transfer devices}

\author{Ya.\ B. Bazaliy}
 \affiliation{Instituut Lorentz, Leiden University, The Netherlands,}
 \affiliation{Department of Physics and Astronomy, University of South Carolina,
 Columbia, SC,}
 \affiliation{Institute of Magnetism, National Academy of Science, Ukraine.}

\date{May, 2006}

\begin{abstract}
Current induced precession states in spin-transfer devices are
studied in the case of large easy plane anisotropy (present in most
experimental setups). It is shown that the effective one-dimensional
planar description provides a simple qualitative understanding of
the emergence and evolution of such states.  Switching boundaries
are found analytically for the collinear device and the spin-flip
transistor. The latter can generate microwave oscillations at zero
external magnetic field without either special functional form of
spin-transfer torque, or ``field-like'' terms, if Gilbert constant
corresponds to the overdamped planar regime.
\end{abstract}


%72.25.Ba Spin polarized transport in metals
%72.25.Hg Electrical injection of spin polarized carriers
%72.25.Mk Spin transport through interfaces
%72.25.Pn Current-driven spin pumping
%75.40.Gb Dynamic properties (dynamic susceptibility, spin waves,
%spin diffusion, dynamic scaling, etc.)
%75.47.De Giant magnetoresistance
%75.60.Ch Domain walls and domain structure (for magnetic bubbles,
%see 75.70.Kw)
%75.70.Cn Magnetic properties of interfaces (multilayers,
%superlattices, heterostructures)
%75.60.Jk Magnetization reversal mechanisms
%85.70.Kh Magnetic thin film devices: magnetic heads (magnetoresistive, inductive, etc.);
%domain-motion devices, etc.
%85.75.-d Magnetoelectronics; spintronics: devices exploiting spin
%polarized transport or integrated magnetic fields
%85.80.Jm Magnetoelectric devices

\maketitle

%\section{Introduction}
Spin-polarized currents are able to change the magnetic
configuration of nanostructures through the spin-transfer effect
proposed more than a decade ago \cite{slon96,berger}. Intensive
research is currently directed at understanding the basic physics of
this non-equilibrium interaction and designing magnetic nanodevices
with all-electric control.

Initial spin-transfer experiments emphasized the current induced
switching between two static configurations \cite{katine2000}.
Presently, the research focus is broadening to include the states
with continuous magnetization precession powered by the energy of
the current source \cite{slon96,sun2000,bjz2004}. Spin-transfer
devices with precession states (PS) serve as nano-generators of
microwave oscillations with remarkable properties, e.g. current
tunable frequency and extremely narrow linewidth
\cite{tsoi2000,kiselev2003,kaka2005,pufall2006}.  A particular issue
of technological importance is the search for systems supporting PS
at zero magnetic field. Here several strategies are pursued: (i)
engineering unusual angle dependence of spin-transfer torque
\cite{manschot2004,gmitra2006,bulle2007}, (ii) relying on the
presence of the ``field-like'' component of the spin torque
\cite{devolder2007}, (iii) choosing the ``magnetic fan'' geometry
\cite{kent2004,lee2005,wang2006,houssameddine2007}.

\begin{figure}[b]
    \resizebox{.45\textwidth}{!}{\includegraphics{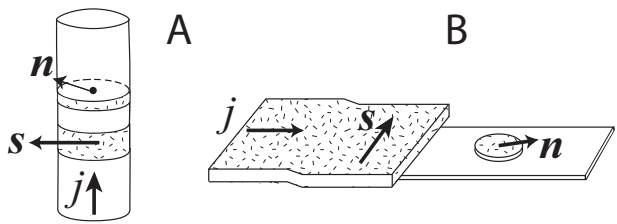}}
\caption{Planar spin-transfer devices. Hashed parts of the devices
are ferromagnetic, white parts are made from a non-magnetic metal.}
 \label{fig:devices}
\end{figure}

PS are more difficult to describe then the fixed equilibria: the
amplitude of precession can be large and non-linear effects are
strong. As a result, information about them if often obtained from
numeric simulations. Here we study PS in planar devices
\cite{bauer-planar-review} using the effective one-dimensional
approximation \cite{weinan-e,boj2007,ybb2007} which is relevant for
the majority of experimental setups. It is shown that planar
approximation provides a very intuitive picture allowing to predict
the emergence of precession and subsequent transformations between
different types of PS. We show that PS in devices with in-plane spin
polarization of the current can exist at zero magnetic field without
the unusual properties (i),(ii) of the spin-transfer torque.

%\section{Generals about the planar effective description}
A conventional spin-transfer device with a fixed polarizer and a
free layer (Fig.~\ref{fig:devices}) is considered. The macrospin
magnetization of the free layer ${\bf M} = M {\bf n}$ has a constant
absolute value $M$ and a direction given by a unit vector ${\bf
n}(t)$. The LLG equation \cite{slon96,bjz2004} reads:
\begin{equation}
 \label{eq:vector_LLG}
{\dot {\bf n}} = \frac{\gamma}{M} \left[ - \frac{\delta E}{\delta
{\bf n}} \times {\bf n} \right] + u({\bf n}) [{\bf n} \times [{\bf
s} \times {\bf n}]] + \alpha [{\bf n} \times \dot {\bf n}] \ .
\end{equation}
Here $\gamma$ is the gyromagnetic ratio, $E({\bf n})$ is the
magnetic energy, $\alpha$ is the Gilbert damping constant, ${\bf s}$
is the spin-polarizer unit vector. The spin transfer strength
$u({\bf n})$ is proportional to the electric current $I$
\cite{bjz2004,boj2007}. In general, it is a function of the angle
between the polarizer and the free layer $u({\bf n}) = f[({\bf
n}\cdot{\bf s})] \ I$, with the function $f[({\bf n}\cdot{\bf s})]$
being material and device specific
\cite{slon2002,kovalev2002,xiao2004}.

\begin{figure}[t]
    \resizebox{.45\textwidth}{!}{\includegraphics{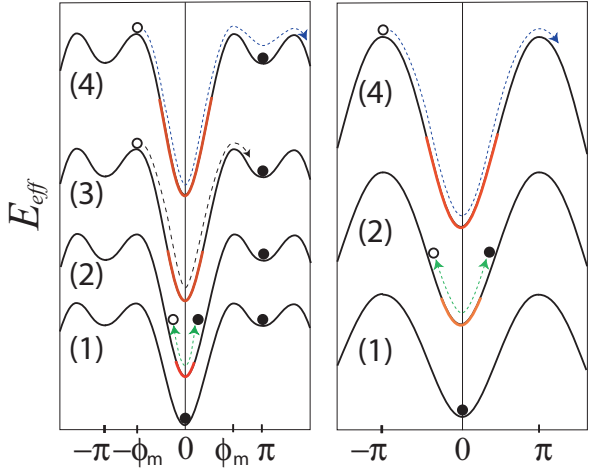}}
\caption{(Color online) Evolution of effective energy profile and
stable solutions with spin-transfer strength (graphs are shifted up
as $u$ becomes more negative) for a device with collinear polarizer.
Left: low-field $0 < h < \tilde\omega_{||}$ regime. Right:
high-field $h
> \tilde\omega_{||}$ regime. Evolution stage (3) is missing in the
high-field regime due to the absence of the second energy minimum.
The red parts of the energy graphs mark the $\alpha_{eff} < 0$
regions. Filled and empty circle gives represent the effective
particle.}
 \label{fig:collinearE}
\end{figure}

The LLG equation can be written in terms of the polar angles
$(\theta,\phi)$ of vector $\bf n$. Planar devices are characterized
by the energy form $E = (K_{\perp}/2)\cos^2\theta +
E_r(\theta,\phi)$ with $K_{\perp} \gg |E_r|$. The first term
provides the dominating easy plane anisotropy and ensures that the
low energy motion happens close to the $\theta = \pi/2$ plane. The
residual energy $E_r$ has an arbitrary form. The smallness of
$\delta\theta = \theta(t) - \pi/2$ allows to derive a single
effective equation on the in-plane angle $\phi(t)$ by performing the
expansion in small parameter $|E_r|/K_{\perp}$ \cite{ybb2007}. For
time-independent current and polarizer direction $\bf s$ one
obtains:
\begin{equation}
 \label{eq:effective_equation}
\frac{1}{\omega_{\perp}} \ddot\phi +  \alpha_{eff}\dot\phi = -
\frac{\gamma}{M} \frac{\partial E_{eff}}{\partial\phi} \ ,
\end{equation}
where $\omega_{\perp} = \gamma K_{\perp}/M$. General expressions for
$\alpha_{eff}(\phi)$ and $E_{eff}(\phi)$ for arbitrary function
$E_r(\theta,\phi)$ and polarizer direction $\bf s$ are given in
Ref.~\onlinecite{ybb2007}. In a special case frequently found in
practice the polarizer ${\bf s}$ is directed in the easy plane at
the angle $\phi_s$, and the residual energy satisfies $(\partial
E_r/\partial\theta)_{\theta = \pi/2} = 0$, i.e. does not shift the
energy minima away from the plane. We will also use the simplest
form $f[({\bf n}\cdot{\bf s})] = {\rm const}$ for the spin transfer
strength. A more realistic function can be employed if needed. With
these assumptions \cite{ybb2007}:
\begin{eqnarray}
\label{eq:alpha_DeltaE_special}
 \alpha_{eff} &=& \alpha +
 \frac{2 u \cos(\phi_s - \phi)}{\omega_{\perp}} \ ,
 \\
 \nonumber
 E_{eff} &=& E_r(\pi/2,\phi)
 - \frac{M u^2}{2\gamma\omega_{\perp}}\cos^2(\phi_s - \phi)  \ .
\end{eqnarray}

\begin{figure}[t]
    \resizebox{.45\textwidth}{!}{\includegraphics{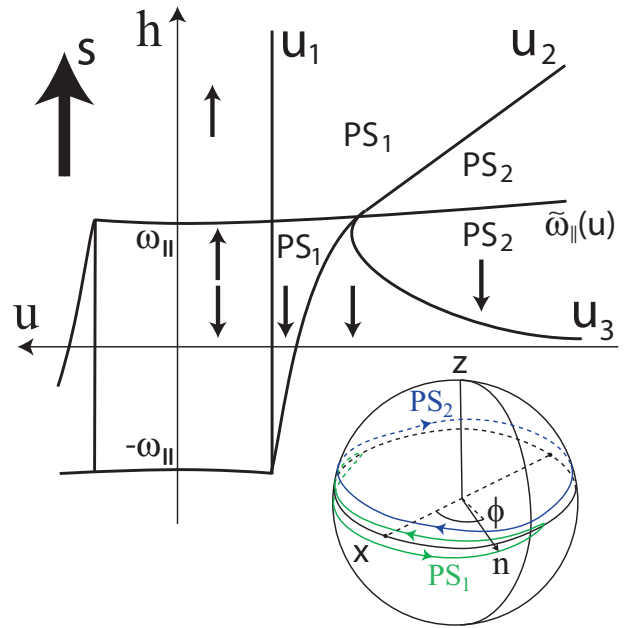}}
\caption{(Color online) Switching diagram of a device with collinear
polarizer. The $u$-axis direction is reversed for the purpose of
comparison with Refs.~\onlinecite{kiselev2003,xiao2005}. The parts
of the diagram not shown can be recovered by a 180-degree rotation
of the picture. Stable directions in each region are given by small
arrows, the precession states are marked as PS$_{1,2}$. The large
arrow shows the polarizer direction. Inset: schematic trajectories
of the PS$_{1,2}$ states on the unit sphere.}
 \label{fig:collinear_switching_diagram}
\end{figure}

Equation (\ref{eq:effective_equation}) has the form of Newton's
equation of motion for a particle in external potential
$E_{eff}(\phi)$ with a variable viscous friction coefficient
$\alpha_{eff}(\phi)$. The advantage of such a description is that
the motion of the effective particle can be qualitatively understood
by applying the usual energy conservation and dissipation arguments.
In the absence of current, the effective friction is a positive
constant, so after an initial transient motion the system always
ends up in one of the minima of $E_r(\pi/2,\phi)$. When current is
present, effective friction and energy are modified. Such a
modification reflects the physical possibility of extracting energy
from the current source, and leads to the emergence of the
qualitatively new dynamic regime of persistent oscillations. These
oscillations of $\phi$ correspond to the motion of $\bf n$ along the
highly elongated ($\delta\theta \ll 1$) closed orbits (see examples
in Fig.~\ref{fig:collinear_switching_diagram}, inset), i.e.
constitute the limiting form of the precession states
\cite{slon96,bjz2004,kiselev2003,xiao2005} in spin-transfer systems.

% \section{Precession states in a collinear device}
To illustrate the advantages of the effective particle description,
consider a specific example of PS in the nanopillar experiment
\cite{kiselev2003} where $E_r$ is an easy axis anisotropy energy
with magnetic field $H$ directed along that axis, $E_r(\phi) =
(K_{||}/2)\sin^2\phi - H M \cos\phi$. The polarizer $\bf s$ is
directed along the same axis with $\phi_s = 0$ (collinear
polarizer). With the definitions $\omega_{||} = \gamma K_a/M$, $h =
\gamma H$, the effective energy becomes \cite{ybb2007}
\begin{equation}\label{eq:Eeffeasyaxis}
\frac{\gamma}{M} E_{eff} = \frac{\tilde\omega_{||}(u)}{2}\sin^2\phi
- h \cos\phi \ ,
\end{equation}
with $\tilde\omega_{||} = \omega_{||} + u^2/\omega_{\perp}$.
Effective energy profiles are shown in Fig.~\ref{fig:collinearE}.
For low fields, $|h| < \tilde\omega_{||}(u)$, the minima at $\phi =
0,\pi$ are separated by maxima at $\pm\phi_m(h)$.

According to Eq.~(\ref{eq:alpha_DeltaE_special}), the effective
friction can become negative at $\phi = 0$ or $\phi = \pi$ at the
critical value of spin-transfer strength $|u| = u_1 =
\alpha\omega_{\perp}/2$. If this value is exceeded, the position of
the system in the energy minimum becomes unstable. Indeed, the
stability of any equilibrium in one dimension depends on whether it
is a minimum or a maximum of $E_{eff}$ and on the sign of
$\alpha_{eff}$ at the equilibrium point. Out of four possible
combinations, only an energy minimum with $\alpha_{eff} > 0$ is
stable. A little above the threshold, $\alpha_{eff}$ is negative in
a small vicinity of the minimum where the system in now
characterized by negative dissipation. In this situation any small
fluctuation away from the equilibrium initiates growing
oscillations. As the oscillations amplitude exceeds the size of the
$\alpha_{eff}<0$ region, part of the cycle starts to happen with
positive dissipation. Eventually the amplitude reaches a value at
which the energy gain during the motion in the $\alpha_{eff} < 0$
region is exactly compensated by the energy loss in the
$\alpha_{eff} > 0$ region: thus a stable cycle solution emerges
(Fig.~\ref{fig:collinearE}, profile (2)).

The requirement of zero total dissipation means that an integral
over the oscillation period satisfies $\int
\alpha_{eff}(\phi)(\dot\phi)^2 dt = 0$. In typical collinear systems
\cite{xiao2005} Gilbert damping satisfies $\alpha \approx 0.01 \ll
\sqrt{\omega_{||}/\omega_{\perp}} \approx 0.1 \ll 1$, hence the
oscillator (\ref{eq:effective_equation}),(\ref{eq:Eeffeasyaxis})
operates in the lightly damped regime. In zeroth order approximation
the friction term in (\ref{eq:effective_equation}) can be neglected,
and a first integral $\dot\phi^2/(2\omega_{\perp}) + E_{eff} = E_0$
exists. Zero dissipation condition can be then approximated by
\begin{equation}\label{eq:dissipative_condition_2}
\int_{\phi_1}^{\phi_2} \alpha_{eff}(\phi) \sqrt{E_0 - E_{eff}(\phi)}
\ d\phi = 0 \ ,
\end{equation}
with $\phi_{1,2}(u)$ being the turning points of the effective
particle trajectory, and $E_0 = E_{eff}(\phi_1) = E_{eff}(\phi_2)$.
Since the integrand of (\ref{eq:dissipative_condition_2}) is a known
function, the formula provides an expression for the precession
amplitude.

Consider now the low positive field regime $0 < h <
\tilde\omega_{||}$. At $u = -u_1$ the parallel configuration becomes
unstable and a cycle emerges near the $\phi = 0$ minimum. As $u$ is
made more negative, the oscillation amplitude grows until eventually
it reaches the point of energy maximum at $u = -u_2$. Equivalently,
the effective particle starting at the energy maximum $-\phi_m$ is
able to reach the other maximum at $+\phi_m$
(Fig.~\ref{fig:collinearE}, left, (3)). Above this threshold the
particle inevitably goes over the potential hill and falls into the
$\phi = \pi$ minimum which remains stable since $\alpha_{eff}(\pi) >
0$ holds for negative $u$. In other words, the cycle solution with
oscillations aroung $\phi = 0$ ceases to exist. At even more
negative $u$ the third threshold is reached when the effective
particle can complete the full rotation starting from the energy
maximum (Fig.~\ref{fig:collinearE}, profile (4)). Below $u = -u_3$ a
new PS with full rotation emerges. In the high-field regime $h
> \tilde\omega_{||}$ the evolution of the precession cycle is
similar (Fig.~\ref{fig:collinearE}, right), but stage (3) is missing
since there is no second minimum. The threshold $u = -u_2$ separates
the finite oscillations regime and the full-rotation regime.

Thresholds $u_i$ can be obtained analytically from
(\ref{eq:dissipative_condition_2}) by substituting the critical
turning points $\phi_{1,2}$ listed above:
\begin{eqnarray}
     \label{eq:u2low}
u_2 &=& \alpha\omega_{\perp} \frac{h \phi_m + \omega_{||}
\sin\phi_m}{\omega_{||} \phi_m + h \sin\phi_m} \quad (h <
\omega_{||}) \ ,
    \\
    \label{eq:u2high}
u_2 &=& \alpha\omega_{\perp} \frac{h}{\omega_{||}} \quad (h >
\omega_{||}) \ ,
    \\
    \label{eq:u3low}
 u_3 &=& \alpha\omega_{\perp} \frac{h (\phi_m-\pi/2) + \omega_{||} \sin\phi_m}
 {\omega_{||} (\phi_m-\pi/2) + h \sin\phi_m} \quad (h < \omega_{||}) \ .
\end{eqnarray}
The corresponding switching diagram is shown in
Fig.~\ref{fig:collinear_switching_diagram} (cf. numerically obtained
Fig.~12 in Ref.~\onlinecite{xiao2005}). It shows that different
hysteresis patterns are possible depending on the trajectory in the
parameter space.

PS in the low field regime was discussed analytically in an
unpublished work \cite{valet2004}. However, since a conventional
description with two polar angles was used, the calculations were
much less transparent. Numeric studies of the PS were performed in
Refs.~\cite{kiselev2003,xiao2005} after the experimental observation
\cite{kiselev2003} of the current induced transition between two PS
in the high field regime. They had shown that indeed the low-current
precession state PS$_1$ has a finite amplitude of
$\phi$-oscillations, while the high-current state PS$_2$ exhibits
full rotations of $\phi$
(Fig.~\ref{fig:collinear_switching_diagram}, inset).

\begin{figure}[t]
    \resizebox{.45\textwidth}{!}{\includegraphics{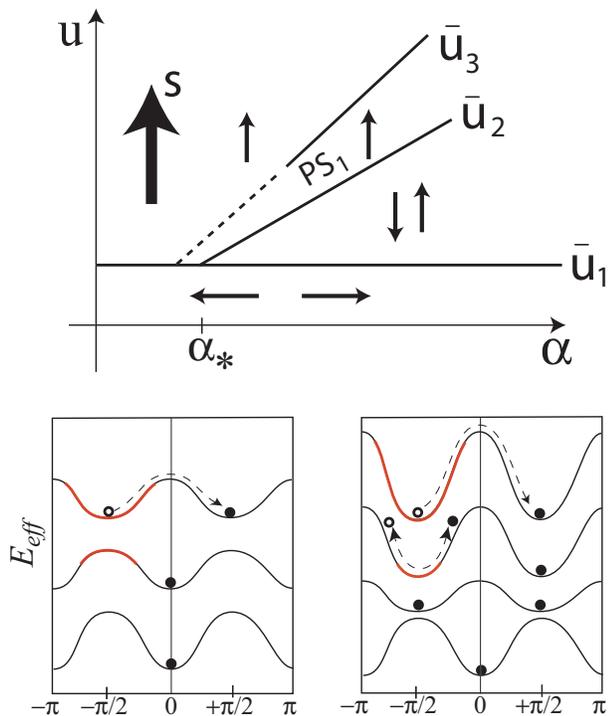}}
\caption{(Color online) Switching diagram of a spin-flip transistor.
The $u < 0$ part of the diagram can be obtained by reflection with
respect to the horizontal axis. In each region stable directions are
given by small arrows, precession state is marked by PS$_{1}$. The
large arrow shows the polarizer direction. Threshold $\bar
u_3(\alpha)$ is sketched as a dashed line where approximation
(\ref{eq:ubar3}) is not valid. Lower panels: the evolution of
effective energy and trajectories (graphs are shifted up with
growing $u$) at $\alpha << \alpha_*$ (left) and $\alpha
> \alpha_*$ (right). The red part of the energy graph marks the
$\alpha_{eff} < 0$ region. Effective particle is shown by filled and
empty circles.}
 \label{fig:sft-cycles}
\end{figure}

%\section{Analysis of the spin-flip transistor}
Next, we consider the cycle solutions in a device called a spin-flip
transistor \cite{bauer-spin-flip-transistor,bauer-planar-review}. It
differs form the setup studied above in the polarizer direction,
which is now perpendicular to the easy axis with $\phi_s = \pi/2$.
No external magnetic field is applied. In this case \cite{ybb2007}
\begin{eqnarray}
 \alpha_{eff} &=& \alpha + \frac{2 u \sin\phi}{\omega_{\perp}} \ ,
\\
 \frac{\gamma}{M}E_{eff} &=& \frac{\bar\omega_{||}(u)}{2} \sin^2\phi
 \ ,
\end{eqnarray}
with $\bar\omega_{||} = \omega_{||} - u^2/\omega_{\perp}$. As the
spin-transfer strength grows, the behavior of the system changes
qualitatively when $\bar\omega_{||}$ or $\alpha_{eff}|_{\pm\pi/2}$
change signs at the thresholds $\bar u_1 =
\pm\sqrt{\omega_{||}\omega_{\perp}}$ and $\bar u_2 = \pm\alpha
\omega_{\perp}/2$. In accord with previous investigations
\cite{morise2005,wang2006} at $|u| > \bar u_1$ the $\phi = 0,\pi$
energy minima are destabilized and the parallel state $\phi = {\rm
sgn}[u]\phi_s$ becomes stable. Surprisingly, for $\alpha
> \alpha_* = 2 \sqrt{\omega_{||}/\omega_{\perp}}$ a window $\bar u_1
< u < \bar u_2$ of stability of antiparallel configuration, $\phi =
-{\rm sgn}[u]\phi_s$, opens (Fig.~\ref{fig:sft-cycles}). As
discussed in Ref.~\cite{ybb2007}, the stabilization of the
antiparallel state happens as the spin-transfer torque is increased
in spite of the fact that this torque repels the system from that
direction. At $u = \bar u_2$ the antiparallel state turns into a
cycle (Fig.~\ref{fig:sft-cycles}, low right panel) which we will
study here. Above the $\bar u_2$ threshold the amplitude of
oscillations grows until they reach the energy maximum at $u = \bar
u_3$ and the cycle solution disappears. Although $\alpha$ is not
small, $\bar u_3$ can still be determined from
Eq.~(\ref{eq:dissipative_condition_2}) because $\alpha_{eff}$ is
small when $u$ is close to $u_2$. Calculating the integral in
(\ref{eq:dissipative_condition_2}) with $\phi_{1,2} = -\pi,0$ we get
\begin{equation} \label{eq:ubar3}
 \bar u_3 = \frac{2}{\pi}\alpha\omega_{\perp} \approx 1.27 \, \bar
 u_2 \ .
\end{equation}
The usage of approximations
(\ref{eq:dissipative_condition_2}),(\ref{eq:ubar3}) is legitimate
for $\alpha \gtrsim 2\alpha_*$ where $\alpha_{eff}(\bar u_3) \ll
\sqrt{\bar\omega_{||}(\bar u_3)/\omega_{\perp}}$ holds. For smaller
values of $\alpha$ numeric calculations are required. They show the
existence of a stable cycle down to $\alpha = 0.8 \, \alpha_*$ where
the stabilization of the antiparallel state is impossible. For
$\alpha \ll \alpha_*$ and $u \gtrsim \bar u_1$ the strong negative
dissipation regime is realized, $|\alpha_{eff}| \gg
\sqrt{\bar\omega_{||}/\omega_{\perp}}$. Numeric results show that
the amplitude of the oscillations induced by negative dissipation is
so big that the effective particle always reaches the energy maximum
and drops into the stable parallel state (Fig~\ref{fig:sft-cycles},
low left panel). We conclude that the line $\bar u_3(\alpha)$
crosses the $u = \bar u_1$ line at some point and terminates there.

As for the full-rotation PS, one can show analytically that it does
not exists in the small dissipation limit at $\alpha \gtrsim 2
\alpha_*$. Numerical simulations do not find it in the $\alpha <
\alpha_*$, $u > \bar u_1$ regime either.

%\section{Conclusions}
In conclusion, we have shown that the planar effective description
can be very useful for studying precession solutions in the spin
transfer systems. It was already used to describe the ``magnetic
fan'' device with current spin polarization perpendicular to the
easy plane \cite{boj2007}. Here the switching diagrams were obtained
for the spin polarizers directed collinearly and perpendicular to
the easy direction within the plane. In collinear case we found
analytic formulas for the earlier numeric results, while the study
of precession solutions in the perpendicular case (spin-flip
transistor) at large damping is new. The latter shows the
possibility of generating microwave oscillations in the absence of
external magnetic field without the need to engineer special angle
dependence of the spin-transfer torque or ``field-like'' terms. The
inequality $\alpha > 2\sqrt{\omega_{||}/\omega_{\perp}}$ required
for the existence of such oscillations can be satisfied by either
reducing the in-plane anisotropy, or increasing $\alpha$ due to
spin-pumping effect \cite{tserkovnyak_RMP}. Most importantly, the
effective planar description allows for qualitative understanding of
the precession cycles and makes it easy to predict their emergence,
subsequent evolution, and transitions between different precession
cycle types. E.g., in the systems with one region of negative
effective dissipation, such as considered here, it shows that no
more then two precession states, one with finite oscillations and
another with full rotations, can exist. Numerical approaches, if
needed, are then based on a firm qualitative foundation. In
addition, numerical calculations in one dimension are easier then in
the conventional description with two polar angles.

%\section{Acknowledgments}
The author thanks G. E. W. Bauer and M. D. Stiles for discussion.
Research at Leiden University was supported by the Dutch Science
Foundation NWO/FOM. Part of this work was performed at Aspen Center
for Physics.

\end{document}